\documentclass[preprint,12pt]{article}
\usepackage{amsfonts,epsfig}
\usepackage{amsmath,amsfonts}
\usepackage{amssymb,graphicx}

\def\pa{\partial}

\def\d{\delta}
\def\h{\hat}
\def\B{\bar}

\def\f{\frac}

\def\p{\varphi}

\def\d{\delta}
\def\h{\hat}

\def\f{\frac}

\def\l{\label}
\def\e{\varepsilon}
\def\a{\alpha}
\def\b{\beta}

\def\g{\gamma}
\def\G{\Gamma}

\def\r{\rho}

\def\S{\Sigma}

\def\be{\begin{equation}}
\def\ee{\end{equation}}
\def\ba{\begin{eqnarray}}
\def\ea{\end{eqnarray}}

\def\r{\rho}

\def\S{\Sigma}
\def\th{\theta}

\def\be{\begin{equation}}
\def\ee{\end{equation}}
\def\ba{\begin{eqnarray}}

\begin{document}

\vspace{5cm}
\centerline{\large\bf
 Black holes with supertranslation field, "large transformations"
 and Israel theorem}
\vspace{1cm}
\centerline{{\bf Mikhail Z. Iofa}
\footnote {e-mail iofa@theory.sinp.msu.ru} }
\centerline{Skobeltsyn Institute of Nuclear Physics}
\centerline{Lomonosov Moscow State University}
\centerline{Moscow, 119991, Russia}

\begin{abstract}

An axial-symmetric vacuum solution of the Einstein equations
containing a supertranslation field
diffeomorphic to the Schwarzschild solution is discussed in the	context	
of Israel theorem.
The metric satisfies all conditions of the Israel theorem, except for
the condition on the form of the metric at spatial infinity. Nevertheless,
following  the steps of the proof of the theorem we show that
the proof applies to the metric with supertranslation field     
and the (transformed) metric used in the proof is spherically symmetric.
We explain the source of the seeming discrepancy connected with the use 
of "large" transformations changing supertranslation field in the metric.

\end{abstract}

\section{Introduction}

The metric of the stationary space-time resulting from the gravitational
collapse of (rotating) black hole is diffeomorphic to the Kerr metric
\cite{carter,robinson,chand}.
Because general diffeomorphisms contain also supertranslations 
(angular-dependent
time translations at the null infinity) \cite{strom1}, the  metric of the
final state of collapse, in general, contains also supertranslation field
associated with supertranslations.

The Israel theorem states that "among all static, asymptotically flat vacuum 
space-times with closed, simply connected equipotential surfaces 
$g_{00}=const$, the Schwarzschild solution is the only one which 
has a nonsingular infinite-red-shift surface $g_{00}=0$" \cite{isr1,isr2}.
Physically, this means that no static
asymmetric perturbation by sources within the horizon can preserve
regularity of the event horizon. Examples of such sources are quadrupole
perturbations \cite{dor},
 magnetic dipole fields \cite{gin}  inside a black hole,  and
small perturbations of the Schwarzschild black hole \cite{RW}.

A family of vacuum solutions of the Einstein equations diffeomorphic to 
Schwarzschild metric and containing a supertranslation field 
was constructed in \cite{comp2}. 
The metric containing a supertranslation field is 
obtained from the Schwarzschild metric
by a diffeomorphism containing a supertranslation field and is
 not spherically symmetric.
The metrics with supertranslation field are physically different 
from the Schwarzschild metric, and
transformations of the metric which depend on supertranslation field   
are not pure (gauge) transformations, and,  in a general case,  
change  superrotation charges
\cite{strom1,comp2,comp1}.

In this note we consider a vacuum axisymmetric solution of the Einstein 
equations with a supertranslation field. 
The metric is obtained by transformation of the solution of \cite{comp2}
and is chosen so that the horizon  of the metric is located at the
sphere of the radius $2M$, where $M$ is the mass parameter of the
solution (and the original Schwarzschild metric).

The metric satisfies all conditions of the Israel theorem, except for
the condition on the form of the metric at spatial infinity. Nevertheless,
repeating  the steps of the proof of the theorem, we show that  
in special coordinates, used in the proof of the theorem, 
the metric is spherically symmetric.

We show that a seeming discrepancy is a consequence of the form of 
transformation to coordinates in which the theorem is proved. 
Transformation from coordinates in which the metric is
initially written to  coordinates used in the proof of the theorem
is not a pure gauge transformation, but a diffeomorphism containing  
supertranslation  field.  
The combined action 
of both "large" transformations - the initial one 
from the Schwarzschild solution to an axisymmetric solution
containing supertranslation field and the second to the preferred
coordinates  used in the proof of the theorem - 
 produces the metric independent of supertranslation field. 

The note is organized as follows. After  short reviews of 
the formulation of the Israel theorem and the form of
 the axial symmetric vacuum solution with supertranslation field,
we discuss the coordinate transformation used in the proof of the theorem.
Next, following the steps of the proof of the theorem, 
we consider application of the inequalities following from finiteness
of the Kretschmann's scalar \cite{isr1,isr2} to the present case.
Applying the inequalities, we conclude that the (transformed) metric is
spherically symmetric and discuss the source of the seeming discrepancy.   
In the
Appendix we  calculate the Kretschmann scalar and
show that it coincides with that for the Schwarzschild solution.

\section{Conditions of the Israel theorem}

The  Israel theorem states that the only vacuum static metric satisfying 
the conditions listed below is the Schwarzschild solution.

The conditions are as follows:
Let $\S$ be a hypersurface $t=const$, maximally extended so that
the square of the Killing vector $\xi$ is negative $\xi\cdot\xi<0$.
It is assumed that the 3D hypersurface $\Sigma$ is regular and
non-compact.
In this case the line element can be locally reduced to a form
\be
\l{n.1}
ds^2 =g_{\a\b}(x^1 ,x^2 ,x^3 )dx^\a dx^\b -V^2 (x^1 ,x^2, x^3 )dt^2
,\ee
where $|V^2|=\xi_\a\xi^\a$ (Greek indices run 1-3) .

The metric has the following asymptotic form 
\ba
\nonumber
1.&{}& g_{\a\b} =\d_{\a \b } +O(r^{-1}),\qquad \pa_\g g_{\a\b}=O(r^{-2}),
\qquad  r^2 =g_{\a\b} x^\a x^\b\rightarrow\infty,\\\nonumber
2.&{}& V=1-M/r +\eta ,\qquad \eta =O(r^{-2}), \qquad \pa_\a \eta = O(r^{-3}),
\qquad \pa_\a\pa_\b\eta = O(r^{-4}),
\ea
and the surfaces $V(x)=const >0$ are  connected closed regular 2D surfaces.
\newline
\hspace*{1.7cm}3.\,\,\,
 4D invariant $R_{ijkl}R^{ijkl}$ is bounded on $\Sigma$.\newline
\hspace*{1.7cm}4. \,\,
If the greatest lower bound of $V$ on $\S$ is zero, then  geometry of
the equipotential surfaces $V=\e$ in the limit $\e\rightarrow 0$ approaches the 
geometry corresponding to a closed regular 2-space of finite area.

Because the metric is a vacuum solution, from the Einstein equations it follows
that
\be
\l{1}
g^{\a\b}R_{\a\b}=0,\qquad R_{\a\b}+ V^{-1}V_{;\a\b}=0,
\ee
$(;)$ denotes covariant derivative. The consequence of (\ref{1}) is that $V$ 
is  harmonic function
\be
\l{2}
g^{\a\b} V_{;\a\b}=0.
\ee
\section{Vacuum solution with supertranslation field}

We consider a class of vacuum, static, asymptotically flat metrics containing 
supertranslation field $C(\th )$ constructed in \cite{comp2}
\be
\l{3}
ds^2 =-\f{(1-M/2\r_s )^2}{(1+M/2\r_s )^2}dt^2 +
(1+M/2\r_s )^4 \left(d\r^2 + (((\r -E)^2 +U)\g_{ab} +(\r -E)C_{ab})dz^a dz^b
 \right).
\ee
The metric (\ref{3}) was obtained from the Schwarzschild metric 
by application of a
diffeomorphism containing a supertranslation field.
 Variables $z^a$ are realized as angles $\th ,\p$
on the unit sphere with the metric
 $ds_{(2)}^2 =\g_{ab}dz^a dz^b=d\th^2 +\sin^2\th d\p^2$.
The functions $C_{ab}$ and  $E,U$ depend on $ C(\th )$ and
its derivatives,
\be
\l{3.1}
\r_s (\r, C) =\sqrt{(\r -C )^2 + {C'}^2(\th ) }.
\ee
The prime is derivative over $\th$.
The horizon of the metric (\ref{3}) is located at 
the surface $\r_s (\r ,C) =M/2$.

By the  transformation
\be
\l{3.2}
r (\r,\th )=\r_s (\r, C)\left(1+\f{M}{2\r_s (\r, C)}\right)^2
,\ee
the metric (\ref{3}) is transformed to a form with the horizon located
at the sphere $r=2M$ \cite{iofa}:
\ba
\nonumber
&{}&ds^2  =-V^2 dt^2 +  \f{dr^2}{V^2(1-b^2 )} +2drd\th\f{br (\sqrt{1-b^2}-b' )}{(1-b^2 )V}+\\
\l{3.4}
&{}& + d\th^2 r^2\f{ (\sqrt{1-b^2}-b' )^2}{(1-b^2 )} +d\p^2 r^2\sin^2\th (b\cot\th
-\sqrt{1-b^2 })^2=\\
\l{4a}
&{}& =-V^2 dt^2 +\f{dr^2 \B{g}_{rr}}{V^2} +\f{2drd\th \B{g}_{r\th}}{V}+
d\th^2 \B{g}_{\th\th}+ d\p^2 \sin^2\th\B{g}_{\p\p}
.\ea
In (\ref{3.4}) we introduced the functions
\be
\l{3.6}
V^2=1-\f{2M}{r},\qquad b=\f{2C'(\th )}{K}, \qquad K=r-M +rV.
\ee
In (\ref{4a}) we separated the powers of $V$ and denoted by
$\B{g}$  the $O(V^0 )$ parts of the metric components.
We  express the  spatial components of the metric and their inverse as
\be
\l{3.7}
\begin{array}{c} {}\\g_{\a\b}=\\{}\\{}\end{array} 
\left|\begin{array}{ccc}
\B{g}_{rr}/{V^2}  & \B{g}_{r\th}/{V}&0\\
 {\B{g}_{r\th}}/{V} & {g}_{\th\th}&0\\
0&0&{g}_{\p\p}
\end{array}\right|
\qquad
\begin{array}{c} {}\\g^{\a\b}=\\{}\\{}\end{array}
\left|\begin{array}{ccc}
V^2  & -{V\B{g}_{r\th}}/{{g}_{\th\th}}&0\\
 -{V\B{g}_{r\th}}/{{g}_{\th\th}} &{\B{g}_{rr}}/{ {g}_{\th\th}}&0\\
0&0&1/{g}_{\p\p}
\end{array}\right|
.\ee
The metric components satisfy the identity
\be
\l{3.8}
{g}_{rr}{g}_{\th\th}-{{g}_{r\th}}^2= \f{{g}_{\th\th}}{V^2}
.\ee
The metric (\ref{3.4}) has the time-like Killing vector
$\xi^i =const \d^i_t$ which becomes null at the horizon.
Solving the geodesic equations for null geodesics, it is possible to show
that the surface $r=2M$ is the surface of infinite redshift \cite{iofa}.

In the limit $r\rightarrow\infty$ the metric (\ref{3.4}) takes a form
\be
\l{3.9}
ds^2 =-dt^2 +dr^2 +2C'(\th )dr d\th + r^2 (d\th^2 +\sin^2\th d\p^2 ).
\ee
The metric satisfies all conditions of the theorem except for
 the first condition
on the form of the metric at spatial infinity.

\section {Transformation to coordinate system used in 
the proof of the theorem}

To prove the theorem, Israel considered a system of the equipotential surfaces
$V=const,\, t=const$ with intrinsic coordinates $\h{x}^A,\,\, A=1,2$. 
Coordinates $\h{x}^A$ are chosen 
to be constant along the trajectories orthogonal to the surfaces $V=const$,
\be
\l{n.2}
g^{\a\b}\pa_\a \h{x}^A\pa_\b V=0
.\ee
In coordinates $x^A ,V$ the spatial part of the metric is
\be
\l{n.2a}
ds^2 =\h{g}_{AB} d\h{x}^A d\h{x}^B +\h{g}_{VV}dV^2.
\ee
Substituting in (\ref{n.2}) the metric components from (\ref{3.7}), we have
\be
\l{n.3}
g^{rr}\pa_r \h{x}^A \pa_r V +g^{r\th}\pa_\th \h{x}^A\pa_r V =0,
\ee
where we used that $\pa_\p V=0$.
The $A=1$ component of Eq. (\ref{n.3}) is
\be
\l{n.5}
V^2 \pa_r \h{x}^1 - V\f{ \B{g}_{r\th}(r,\th ) }{{g}_{\th\th}(r,\th )} 
\pa_\th \h{x}^1 =0
.\ee
For the coordinate $\h{x}^2$ we take coordinate $\p$. A general solution of the 
partial differential equation  (\ref{n.5}) is 
$\h{x}^1 =F(\psi (r,\th ))$, where $F$
is an arbitrary function and $\psi (r,\th )= const$ is the integral 
of the ordinary differential equation
\be
\l{n.6}
dr  =-\f{ d\th V {g}_{\th\th}(r,\th ) }{\B{g}_{r\th}(r,\th ) }
.\ee
From Eqs.(\ref{n.2})-(\ref{n.3}) it follows that
\be
\l{n.7}
\h{g}^{AV} =\h{g}_{AV}=0.
\ee
Using the explicit form of $g^{\a\b}$ [Eq. (\ref{3.7})], we have
\ba
\nonumber
&{}&\h{g}^{VV}=g^{\a\b}\pa_\a V\pa_\b V = g^{rr}(\pa_r V)^2= (V\pa_r V )^2\\
&{}&
\l{n.8}
\h{g}^{12}=g^{\a\b}\pa_\a \h{x}^1\pa_\b \h{x}^2 =
g^{\th\p}\pa_\th \h{x}^1 (\th ,r)\pa_\p \p 
+g^{r\p}\pa_r \h{x}^1 (\th ,r)\pa_\p \p =0, 
\ea
where it was used that $g^{\th\p}=g^{r\p}=0$. In coordinates $(x^A ,V)$
the metric is
\be
\l{n.9}
d\h{s}^2 = \h{g}_{VV}dV^2 + \h{g}_{11}d\h{x}^1 d\h{x}^1 
+\h{g}_{22}d\h{x}^2 d\h{x}^2.
\ee
Equating  the expressions for the volume elements in  coordinates $(r,\th,\p)$
and $(x^A ,V)$ [Eq. (\ref{n.9})], we have
$$
dr d\th d\p\sqrt{det(g)}=dV d^2\h{x} \sqrt{det(\h{g})}
.$$
Using Eqs.(\ref{3.7}), \ref{3.8}), we obtain the volume element in coordinates 
$(r,\th,\p)$ as
\be
dr d\th	d\p \sqrt{ (g_{rr}g_{\th\th}-g^2_{r\th})g_{\p\p} }=
dr d\th d\p\sqrt{ {g}_{\p\p}{g}_{\th\th}V^{-2} }
.\ee
In coordinates $(x^A ,V)$ the volume element is
\ba
dV d^2\h{x}\sqrt{ \h{g}_{11}\h{g}_{22}\h{g}_{VV} }=dr \f{\pa V}{\pa r} d^2\h{x} 
\sqrt{\h{g}_{11}\h{g}_{22}(V\pa_r V)^{-2} }
,\ea
and thus
\be
\l{n.11}
 d\th d\p\sqrt{ g_{\th\th}g_{\p\p} }=d^2 \h{x} \sqrt{\h{g}_{11}\h{g}_{22}V^{-2} }.
\ee

\section{Israel inequalities}

The proof of the theorem is based on the use of relations which follow from 
the assumptions of the theorem, notably from finiteness of the Kretschmann's 
scalar and harmonic property of $V$ (\ref{2}).

The expressions for the extrinsic curvature of the surface 
$V=const$ and related formulas are
\be
\l{n.13}
K_{AB} = \f{1}{2\r}\f{\pa \h{g}_{AB}}{\pa V},\qquad
K=K_{AB}= K_{AB}\h{g}^{AB},\qquad
\f{\pa\sqrt{\h{g}}}{\pa V}=\sqrt{\h{g}}\r K, \qquad
\ee
From (\ref{n.13}) and (\ref{2}) follows the equation
\be
\l{n.14}
\f{\pa}{\pa V} \left(\f{ \sqrt{\hat{g}}}{\r }\right)=0
,\ee
where $$\h{g} =\h{g}_{11}\h{g}_{22}, \qquad 
\r (r) =(g^{rr}\pa_r V \pa_r V )^{-1/2}=(V \pa_r V )^{-1}. $$ 

Integrating (\ref{n.14}) over $\S$, we have
\be
\l{n.15}
0= \int dV d^2\h{x}\f{\pa}{\pa V} \left(\f{\sqrt{\h{g} }}{\r }\right)=
\int_{V\rightarrow 1}d^2 \h{x}\f{\sqrt{\h{g}}}{\r}-
\int_{V\rightarrow 0}d^2 \h{x}\f{\sqrt{\h{g}}}{\r}
=4\pi -S_0/\r_0
.\ee
Here we introduced
$$
S_0 =\int_{V\rightarrow 0}d^2\h{x}\sqrt{\h{g}},
\qquad \r_0 = \lim\limits_{V\to 0} \r (r)=const
$$
and
$$
\int_{V\rightarrow 1}d^2 \h{x}\f{\sqrt{\h{g}}}{\r}=
\int_{V\rightarrow 1}d\th d\p\f{\sqrt{g}}{\r}=
\int_{r\rightarrow\infty}d\th d\p \f{\sqrt{ {g}_{\th\th}{g}_{\p\p} } }{r^2}=4\pi
.$$
Here we used that from (\ref{3.4})  it follows that in the limit $r\rightarrow\infty$
$$
{g}_{\th\th}=r^2 (1+O(r^{-1}),\qquad {g}_{\p\p}=r^2 \sin^2\th (1+O(r^{-1}).
$$
In \cite{isr1} it was shown that in a general case the relation 
$\r_0=\r (0_+,\h{x})=const$ 
follows from condition 3 of the theorem  
that the Kretschmann scalar 
\be
\l{16}
\f{1}{8}R_{ijkl}R^{ijkl}=\f{1}{(V\r )^2}\left[K_{AB}K^{AB} +2\r^{-2}\r_{;A}\r^{;A}+
\r^{-4}(\pa\r/\pa V )^2 \right]
\ee
is bounded in $\S $.

Next, in \cite{isr1,isr2} were obtaind the identities 
\ba
\l{17}
\f{\pa}{\pa V}\left(\sqrt{\f{\h{g} }{ \r } }\f{K}{V}\right)
=-\f{ \sqrt{\h{g} }}{V}\left[ \nabla^2 (\r ^{1/2})
+\f{1}{2} \r^{-3/2} \r_{;A}\r^{;A} + \r^{1/2}(K_{AB}K^{AB}-K^2 /2)\right]\\
\l{18}
\f{\pa}{\pa V}\left[\f{\sqrt{ \h{g} }}{\r }\left( KV +\f{4}{\r}\right)\right]=
-\sqrt{\h{g} }V\left[\nabla^2(\ln\r ) +\r_{;A}\r^{;A} 
+2K_{AB}K^{AB}-K^2 -\h{R}^{(2)}\right]
.\ea
Integrating the first identity over $\S$ and noting that the r.h.s. of (\ref{17}) 
not positive,
one obtains the inequality
\be
\l{19}
\int d^2\h{x}dV
\f{\pa}{\pa V}\left(\sqrt{\f{\h{g} }{ \r }}\f{K}{V}\right)=
\int\limits_{V\rightarrow 1}d^2 x\sqrt{g}\sqrt{\f{K}{\r V}}-
\int\limits_{V\rightarrow 0}d^2 \h{x}\sqrt{\h{g}}\sqrt{\f{K}{\r V}}\leq 0
\ee
In the limit $r\rightarrow\infty \quad K\simeq 2/r,\,\, \r\simeq r^2/M$.  
The first integral
in the rhs of (\ref{19}) is $8\pi \sqrt{M}$. The second integral, using the relation
$\lim\limits_{V \to 0} K/V=-(1/2)\r_0 R^{(2)}(0_+, \h{x})$ 
which follows from Eq.(\ref{16}), 
and 
$$\int_\S d^2\h{x} \sqrt{\h{g}}\h{R}^{(2)}=-8\pi ,                                        $$
    is equal to $4\pi \sqrt{\r_0}$. It follows that
\be
\l{20}
2\sqrt{M}-\sqrt{\r_0}\leq 0.
\ee
The integration of Eq.(\ref{18}) over $\S$ yields
\ba
\l{21}\nonumber
&{}&\int\limits_{V\rightarrow 1}d^2\h{x}\sqrt{ {g}}\f{1}{\r}\left(K+\f{4}{\r}\right)
-\int\limits_{V\rightarrow 0}d^2\h{x}\sqrt{ \h{g}}\f{4}{\r_0^2}=\\
&{}&
=-\int d^2\h{x}\sqrt{ \h{g}}\,\int_0^1 dV V\left[\nabla^2(\ln\r ) +
\r_{;A}\r^{;A} +2K_{AB}K^{AB}-K^2 \right]-4\pi
,\ea
where it was used that
$$
\int d^2\h{x}\sqrt{ \h{g}}\int_0^1\,dV V \h{R}^{(2)}=-4\pi
.$$
The integral in the rhs of Eq.(\ref{21}) is not positive,
the first integral in lhs vanishes leaving the inequality
\be
\l{22}
4\pi\leq\f{4 S_0}{\r_0^2}
.\ee
Repeating the reasoning of \cite{isr1,isr2}, from the (in)equalities 
(\ref{n.15}), (\ref{20}), and (\ref{22}) and harmonic property (\ref{3}), 
we conclude that the metric 
Eq.(\ref{n.9}) is spherically symmetric, i.e. is 
the Schwarzschild metric.

The initial  metric  (\ref{3.4}) contained 
a supertranslation field and was axisymmetric.
The metric (\ref{3.4}) was obtained 
from the Schwarzschid metric by application of a diffeomorphism containing
a supertranslation field.
Diffeomorphism containing supertranslation field is 
not a pure gauge transformation (transformation conserving
all the charges), but changes the superrotation charge \cite{strom1,comp1,comp2}. 

Transformation from coordinates $x^\a$ in the metric (\ref{3.4})
to coordinates $\h{x}^A$ again includes
supertranslation field (\ref{n.5}) and is not a pure gauge transformation.
The actions of both diffeomorphisms, the first from the Schwarzshild metric to  
Eq.(\ref{3.4}) and the second from Eq.(\ref{3.4}) to Eq.(\ref{n.9})
result in the metric without supertranslation field, and we returned to the
Schwarzschild metric .

\vspace{1cm}
{\large\bf Acknowledgments}

This work was partially supported by the Ministry of Science and 
Higher Education of
Russian Federation under the project 01201255504.

\section{Appendix:  calculation of the Kretschmann scalar}
\setcounter{equation}{0}
\renewcommand{\theequation}{A\arabic{equation}}

In this Appendix we calculate the Kretschmann scalar for the metric
(\ref{3.4}). The Kretchmann scalar can be presented in a form \cite{isr1,muller}
\be
\l{1a}
\f{1}{8}R_{ijkl}R^{ijkl}= V^{-2}V_{;\a\b}V^{;\a\b}
.\ee
Here the Latin indices run over $0-3$,the  Greek ones over $1-3$.

Calculating $V_{;\a\b}$, we obtain
\ba
\nonumber
&{}&V_{;rr}=V_{,rr}-\G^r_{rr}V_{,r}
=-\f{M(2rV^2  +M)}{r^4 V^3 }-
\f{1}{2}\left[{g}^{rr}{g}_{rr,r}+
{g}^{r\th}(2 {g}_{r\th ,r}-{g}_{rr,\th})\right]V_{,r}=
\\ \l{3a}
&{}&=-\f{2M}{Vr^3 }-\f{M}{2Vr^2}
\left( \B{g}_{rr ,r}-2\f{ \B{g}_{r\th} \B{g}_{r\th,r} }{\B{g}_{\th\th} }
+\f{ \B{g}_{rr,\th} \B{g}_{r\th} }{V\B{g}_{\th\th} }\right),
\\ \l{5a}
&{}&V_{;r\th}=-\G^r_{r\th}V_{,r}=-\f{M}{2Vr^2}\left(\f{1}{2}g^{rr}g_{rr,\th}
+\f{1}{2} g^{r\th}g_{\th\th,r}\right),
\\  \l{6a}
&{}&V_{;\th\th}=-\G^r_{\th\th}V_{,r}=-\f{M}{Vr^2}\left(\f{1}{2}g^{rr}
(2 g_{r\th} -g_{\th\th,r})+\f{1}{2}g^{r\th}g_{\th\th,\th}\right)
 ,\\   \l{7a}
&{}&V_{;\p\p}=-\G^r_{\p\p}V_{,r}=-\f{M}{Vr^2}
\left(\f{1}{2}g^{rr}g_{\p\p,r}+\f{1}{2}g^{r\th}g_{\p\p,\th}\right)
,\\  \l{8a}
&{}&V_{;r\p}=V_{;\th\p}=0.
\ea
Using Eqs.(\ref{2}) and (\ref{3.8}), we rearrange the terms in 
$V_{;\a\b}V^{;\a\b}$ and present it in a form
\be
\l{9a}
V_{;\a\b}V^{;\a\b}=2(V_{;rr}V_{;\th\th}-V^2_{;r\th})(({g}^{r\th})^2 -
{g}^{rr}{g}^{\th\th})+2(V_{;\p\p}{{g}^{\p\p})^2 }
.\ee
Using Eq.(\ref{3.8}), we have
\be
\l{10a}
({g}^{r\th})^2 -
{g}^{rr}{g}^{\th\th} =
V^2\f{\B{g}_{r\th}^2 -\B{g}_{rr}{g}_{\th\th}}{{g}^2_{\th\th} }=
-\f{V^2}{{g}_{\th\th}}.
\ee
Introducing the notations
$$
f= \sqrt{1-b^2 }-b' ,\qquad N= 1-b^2 
,$$
and noting the relations
$$
K_{,r}=K/rV , \qquad b_{,r}=-b/rV ,\qquad V_{,r}=M/Vr^2
,$$
where $b$ and $K$ were defined in  (\ref{3.6}),
we calculate Eqs.(\ref{3a}) -(\ref{6a}) as
\ba
\l{11a} 
&{}&V_{;rr}=V_{,r}\left(-\f{2}{r}+\f{b^2}{rN}\right) ,\\
\l{12a}
&{}&V_{;\th\th}=V_{,r}\f{V^2 r f^2}{N},\\
\l{13a}
&{}&V_{;r\th}=V_{,r}\f{Vbf}{N},\\
\l{14a}
&{}&V_{;\p\p}=V_{,r}\f{V^2 g_{\p\p}}{r}.
\ea
Substituting these expressions into Eq.(\ref{9a}), we finally obtain
\be
\l{15a}
V^{-2}V_{;\a\b}V^{;\a\b}=\f{6M^2}{r^6}.
\ee
The expression (\ref{15a}) coincides with the Kretschmann scalar for the 
Schwarzschild
metric, because the metric with supertranslation field Eq.(\ref{3.8}) and the 
Schwarzschild metric are connected by a diffeomorphism.

\end{document}